\documentclass[11pt]{article}

\usepackage{times, amssymb, fullpage}

\newtheorem{fact}{Fact}    
\newtheorem{result}{Result}    
\newtheorem{definition}{Definition} 
\newtheorem{proposition}{Proposition}    
\newtheorem{theorem}{Theorem}    
\newtheorem{corollary}{Corollary}    
\newtheorem{lemma}{Lemma}      
\newcommand{\qed}{\hfill{$\rule{6pt}{6pt}$}} 
\newenvironment{proof}{\noindent{\bf Proof}:}{\qed}

\newcommand{\defeq}{\stackrel{\Delta}{=}}
\newcommand{\ket}[1]{| #1 \rangle}
\newcommand{\ketbra}[1]{| #1 \rangle \langle #1 |}
\newcommand{\braket}[2]{\langle #1 | #2 \rangle}
\newcommand{\unibraket}[3]{\langle #1 | #2 | #3 \rangle}
\newcommand{\tuple}[1]{\langle #1 \rangle}

\newcommand{\Tr}{\mbox{{\rm Tr} }}
\newcommand{\totvar}[1]{\left\| #1 \right\|_1}
\newcommand{\norm}[1]{\|{ #1 }\|}

\newcommand{\event}{{\cal E}}
\newcommand{\IC}{\mathrm{IC}}
\newcommand{\EQ}{\mathrm{EQ}}
\newcommand{\linspan}{{\mathrm{Span}}}
\newcommand{\range}{\mathsf{range}}

\newcommand{\nat}{{\mathbb N}_+}
\newcommand{\R}{{\mathbb R}}
\newcommand{\C}{{\mathbb C}}
\newcommand{\E}{\mathop{\rm E}}

\newcommand{\cM}{{\cal M}}

\newcommand{\bO}{{\mathbf O}}
\newcommand{\bU}{{\mathbf U}}

\newcommand{\bGamma}{{\mathbf \Gamma}}
\newcommand{\bmu}{{\mathbf \mu}}
\newcommand{\Good}{{\mathsf{Good}}}
\newcommand{\Bad}{{\mathsf{Bad}}}

\newcommand{\cX}{{\cal X}}
\newcommand{\cY}{{\cal Y}}
\newcommand{\cZ}{{\cal Z}}
\newcommand{\bY}{{\mathbf Y}}
\newcommand{\by}{{\mathbf y}}
\newcommand{\bX}{{\mathbf X}}
\newcommand{\bx}{{\mathbf x}}
\newcommand{\bZ}{{\mathbf Z}}

\title{A direct sum theorem in communication complexity via 
message compression}

\author{
Rahul Jain\thanks{
School of Technology and Computer Science,
Tata Institute of Fundamental Research,
Mumbai 400005,
India.
Email: {\sf rahulj@tcs.tifr.res.in}.
}
\and
Jaikumar Radhakrishnan\thanks{
School of Technology and Computer Science,
Tata Institute of Fundamental Research,
Mumbai 400005,
India.
Email: {\sf jaikumar@tcs.tifr.res.in}.
Part of this work was done while visiting MSRI, Berkeley.
}
\and
Pranab Sen\thanks{
Department of Combinatorics and Optimization,
University of Waterloo,
Waterloo, Ontario N2L 3C1,
Canada.
Email: {\sf p2sen@cacr.math.uwaterloo.ca}. 
This work was done while visiting TIFR, Mumbai and MSRI, Berkeley. 
}
}

\date{}

\begin{document}

\maketitle

\begin{abstract} 
We prove lower bounds for the {\em direct sum} problem for two-party
bounded error randomised multiple-round communication protocols.  Our
proofs use the notion of {\em information cost} of a protocol, as
defined by Chakrabarti et al.~\cite{chakra:simul} and refined further
by Bar-Yossef et al.~\cite{bar:ic}.  Our main technical result is a
`compression' theorem saying that, for any probability distribution
$\mu$ over the inputs, a $k$-round private coin bounded error protocol
for a function $f$ with information cost $c$ can be converted into a
$k$-round deterministic protocol for $f$ with bounded distributional
error and communication cost $O(kc)$.  We prove this result using a
{\em substate} theorem about {\em relative entropy} and a {\em
rejection sampling} argument. Our direct sum result follows from this
`compression' result via elementary information theoretic arguments.

We also consider the direct sum problem in quantum communication.
Using a probabilistic argument, we show that messages cannot be
compressed in this manner even if they carry small information.
Hence, new techniques may be necessary to tackle the direct sum
problem in quantum communication.
\end{abstract}

\section{Introduction}
We consider the two-party {\em communication complexity}
of computing a function $f: \cX \times \cY \rightarrow
\cZ$. There are two players Alice and Bob. Alice is given an input
$x \in \cX$ and Bob is given an input $y \in \cY$.  They then
exchange messages in order to determine $f(x,y)$.  The goal is to
devise a protocol that minimises the amount of communication. In the
{\em randomised} communication complexity model, Alice and Bob are
allowed to toss coins and base their actions on the outcome of these
coin tosses, and are required to determine the correct value with high
probability for every input. There are two models for randomised
protocols: in the {\em private coin} model the coin tosses are
private to each player; in the {\em public coin} model the two
players share a string that is generated randomly (independently of
the input). A protocol where $k$ messages are exchanged between the
two players is called a $k$-round protocol. One also considers
protocols where the two parties send a message each to a referee who
determines the answer: this is the {\em simultaneous} message model.

The starting point of our work is a recent result of Chakrabarti,
Shi, Wirth and Yao~\cite{chakra:simul} concerning the {\em direct sum}
problem in communication complexity. For a function $f: \cX \times \cY
\rightarrow \cZ$, the $m$-fold {\em direct sum} is the function
$f^m: \cX^m \times \cY^m \rightarrow \cZ^m$, defined by
$f^m(\tuple{x_1, \ldots, x_m}, \tuple{y_1, \ldots, y_m}) \defeq
\tuple{f(x_1, y_1), \ldots, f(x_m, y_m)}$. One then studies the
communication complexity of $f^m$ as the parameter
$m$ increases. Chakrabarti et
al.~\cite{chakra:simul} considered the direct sum problem in the
bounded error simultaneous message private coin model and showed that 
for the equality function $\EQ_n: \{0,1\}^n
\times \{0,1\}^n \rightarrow \{0,1\}$, the 
communication complexity of $\EQ_n^m$ is $\Omega(m)$ times the 
communication complexity of
$\EQ_n$. In fact, their result is more general. Let $R^{{\rm sim}}(f)$
be the bounded error simultaneous message
private coin communication complexity of $f:\{0,1\}^n \times \{0,1\}^n
\rightarrow \{0,1\}$, and let 
$\tilde{R}^{{\rm sim}}(f)\defeq\min_{S} R^{{\rm sim}}(f|_{S\times S})$, 
where $S$ ranges
over all subsets of $\{0,1\}^n$ of size at least $(\frac{2}{3}) 2^n$.

\paragraph{Theorem (\cite{chakra:simul})} 
$R^{{\rm sim}}(f^m)= \Omega(m(\tilde{R}^{{\rm sim}}(f)-O(\log n)))$. 
A similar result holds for two-party bounded error one-round 
protocols too. 

\bigskip

The proof of this result in~\cite{chakra:simul} had two parts. The first
part used the notion of information cost of randomised protocols,
which is the mutual information between the inputs (which were chosen
with uniform distribution in~\cite{chakra:simul}) and the 
transcript of the
communication between the two parties.  Clearly, the information cost
is bounded by the length of the transcript. So, showing lower bounds
on the information cost gives a lower bound on the communication
complexity. Chakrabarti et al.{} showed that the information cost is
super-additive, that is, the information cost of $f^m$ is at
least $m$ times the
information cost of $f$. The second part of their argument showed an
interesting message compression result for communication protocols.
This result can be stated informally as follows: if the message
contains at most $a$ bits of information about a player's input, then
one can modify the (one-round or simultaneous message) protocol so
that the length of the message is $O(a+\log n)$. Thus, one obtains a
lower bound on the information cost of $f$ if one has a suitable lower
bound on the communication complexity $f$. By combining this with the
first part, we see that the communication complexity of $f^m$ is at
least $m$ times this lower bound on the communication complexity of
$f$. 

In this paper, we examine if this approach can be employed for
protocols with more than one-round of communication. Let
$R^{k}_\delta(f)$ denote the $k$-round private coin communication
complexity of $f$ where the protocol is allowed to err with
probability at most $\delta$ on any input. Let $\mu$ be a probability
distribution on the inputs of $f$.  Let $C^k_{\mu,
\delta}(f)$ denote the deterministic $k$-round communication
complexity of $f$, where the protocol errs for at most $\delta$
fraction, according to the distribution $\mu$, of the inputs.  Let
$C^{k}_{[\,], \delta}(f)$ denote the maximum, over all product
distributions $\mu$, of $C^k_{\mu, \delta}(f)$. We prove the following.

\paragraph{Theorem:}
Let $m, k$ be positive integers, and $\epsilon, \delta > 0$. Let
$f: \cX \times \cY \rightarrow \cZ$ be a function. Then,
\( \textstyle
R^k_\delta(f^m) \geq m \cdot (\frac{\epsilon^2}{2k} \cdot
                                   C^{k}_{[\,], \delta+2\epsilon}(f)
				   - 2
			     ).
\) 

\bigskip

\noindent The proof this result, like the proof 
in~\cite{chakra:simul}, has
two parts, where the first part uses a notion of 
information cost for $k$-round protocols, and the second shows how
messages can be compressed in protocols with low information cost. 
We now informally describe the ideas behind these results.
To keep our presentation simple, we will assume that
Alice's and Bob's inputs are chosen uniformly at random from their
input sets. 

The first part of our argument uses the extension of the notion of
information cost to $k$-round protocols. The information cost of a
$k$-round randomised protocol is the mutual information between the
inputs and the transcript. This natural extension, and its refinement
to {\em conditional information cost} by~\cite{bar:ic} has proved
fruitful in several other contexts~\cite{bar:ic,jrs:disjointness}. It
is easy to see that it is bounded above by the length of the
transcript, and a lower bound on the information cost of protocols
gives a lower bound on the randomised communication complexity. The
first part of the argument in~\cite{chakra:simul} is still applicable:
the information cost is super-additive; in particular, the $k$-round
information cost of $f^m$ is at least
$m$ times the $k$-round information cost of $f$.

The main contribution of this work is in the second part of the
argument. This part of Chakrabarti et al.~\cite{chakra:simul} used a
technical argument to compress messages by exploiting the fact that
they carry low information. Our proof is based on the connection
between mutual information of random variables and the relative
entropy of probability distributions (see Section~\ref{sec:prelim} for
definition). Intuitively, it is reasonable to expect that if the
message sent by Alice contains little information about her input $X$,
then for various values $x$ of $X$, the conditional distribution on
the message, denoted by $P_x$, are similar.  In fact, if we use
relative entropy to compare distributions, then one can show that the
mutual information is the average taken over $x$ of the relative
entropy $S(P_x \| Q)$ of $P_x$ and $Q$, where $Q=\E_X[P_X]$. Thus, if
the information between Alice's input and her message is bounded by
$a$, then typically $S(P_x \| Q)$ is about $a$. To exploit this fact,
we use the Substate theorem of~\cite{jain:substate} which states
(roughly) that if $S(P_x
\| Q) \leq a$, then $P_x \leq 2^{-a} Q$. Using a standard {\em
rejection sampling} idea we then show that Alice can restrict herself
to a set of just  $2^{O(a)} n$ messages; consequently, her messages can
be encoded in $O(a + \log n)$ bits. In fact, such a compact set of
messages can be obtained by sampling $2^{O(a)} n$ times from
distribution $Q$. 

We believe this connection between relative
entropy and sampling is an important contribution of this work.
Besides giving a more direct proof of the second part of
Chakrabarti et al.'s~\cite{chakra:simul} argument,
our approach quickly generalises to two party bounded error
private coin multiple round protocols,
and allows us to prove a message compression result and a
direct sum lower bound for such protocols. Direct sum lower bounds for
such protocols were not known earlier. In addition, our message
compression result and direct sum lower bound
for multiple round protocols 
hold for protocols computing relations too.

The second part of our argument raises an interesting question in the
setting of quantum communication. Can we always make the length of
quantum messages comparable to the amount of information they carry
about the inputs without significantly changing the error probability
of the protocol? That is, for $x \in \{0, 1\}^n$, instead of
distributions $P_x$ we have density matrices $\rho_x$ so that the
expected quantum relative entropy $\E_X[S(\rho_x \| \rho)]
\leq a$, where $\rho \defeq \E_X[\rho_x]$.  Also, we are given
measurements (POVM elements) $M^x_y$, $x, y \in \{0, 1\}^n$.  Then, we
wish to replace $\rho_x$ by $\rho'_x$ so that there is a subspace of
dimension $n \cdot 2^{O(a/\epsilon)}$ that contains the support of
each $\rho'_x$; also, there is a set $A \subseteq \{0, 1\}^n$, $|A|
\geq \frac{2}{3} \cdot 2^n$ such that for each $(x, y) \in A \times
\{0, 1\}^n$, $|\Tr M^x_y \rho_x - \Tr M^x_y \rho'_x| \leq \epsilon$.
Fortunately, the quantum analogue of the Substate theorem has already
been proved by Jain, Radhakrishnan and Sen~\cite{jain:substate}.
Unfortunately, it is the rejection sampling argument that does not
generalise to the quantum setting. Indeed, we can prove the following
strong negative result about compressibility of quantum information:
For sufficiently large constant $a$, there exist $\rho_x$, $M^x_y$,
$x, y \in \{0, 1\}^n$ as above such that any subspace containing the
supports of $\rho'_x$ as above has dimension at least $2^{n/6}$.  This
strong negative result seems to suggest that new techniques may be
required to tackle the direct sum problem for quantum communication.

\subsection{Previous results}
The direct sum problem for communication complexity has been
extensively studied in the past 
(see Kushilevitz and Nisan~\cite{kn:cc}).
Let $f: \{0,1\}^n \times \{0,1\}^n \rightarrow \{0,1\}$ be a function.
Let $C(f)$ ($R(f)$) denote the deterministic (bounded error private
coin randomised) two-party communication complexity of $f$.
Ceder, Kushilevitz, Naor and
Nisan~\cite{FKNN:amortized} showed that there exists a partial
function $f$ with $C(f) = \Theta(\log n)$, whereas solving $m$ copies
takes only $C(f^m) = O(m + \log m \cdot
\log n )$. They also showed a lower bound $C(f^m) \geq m(
\sqrt{C(f)/2} -\log n -O(1))$ for total functions $f$. For the
one-round deterministic model, they showed that 
$C(f^m) \geq m (C(f) - \log n - O(1))$ even for partial functions.
For the two-round deterministic model, Karchmer,
Kushilevitz and Nisan~\cite{KKN92:fractionalcover} showed that 
$C(f^m) \geq m (C(f) -O(\log n))$ for any relation $f$.
Feder et al.~\cite{FKNN:amortized} also
showed that for the equality problem 
$R(EQ^m_n) = O(m + \log n )$.

\subsection {Our results} 
\label{sec:results}
We now state the new results in this paper.

\begin{result}[Compression result, multiple-rounds]
Suppose that $\Pi$ is a $k$-round private coin randomised protocol for 
$f: \cX \times \cY \rightarrow \cZ$. Let the average
error of $\Pi$ under a probability distribution $\mu$ on
the inputs $\cX \times \cY$ be
$\delta$. Let $X, Y$ denote the random variables corresponding to
Alice's and Bob's inputs respectively.
Let $T$ denote the complete transcript of 
messages sent by Alice
and Bob. Suppose $I(XY:T) \leq a$. Let $\epsilon > 0$.
Then, there is another deterministic protocol $\Pi'$
with the following properties: 
\begin{enumerate} 
\item[(a)] The communication cost of $\Pi'$
is at most 
$\frac{2k(a+1)}{\epsilon^2} + \frac{2k}{\epsilon}$ bits;
\item[(b)] The distributional  error  of $\Pi'$ 
under $\mu$ is at most $\delta + 2\epsilon$.
\end{enumerate}
\end{result}

\begin{result}[Direct sum, multiple-rounds] 
Let $m, k$ be positive integers, and $\epsilon, \delta > 0$. Let
$f: \cX \times \cY \rightarrow \cZ$ be a function. Then,
\(
R^k_\delta(f^m) \geq m \cdot \left(\frac{\epsilon^2}{2k} \cdot
                                   C^{k}_{[\,], \delta+2\epsilon}(f)
				   - 2
			     \right).
\)
\end{result}
\begin{result}[Quantum incompressibility]
Let $m, n, d$ be positive integers and $k \geq 7$. Let $d \geq 160^2$,
$1600 \cdot d^4 \cdot k 2^k \ln(20 d^2) < m$ and
$3200 \cdot d^5 \cdot 2^{2k} \ln d < n$.
Let the underlying Hilbert space be $\C^m$. There exist $n$
states $\rho_l$ and $n$ orthogonal projections $M_l$,
$1 \leq l \leq n$, such that
\begin{enumerate}
\item[(a)] $\forall l \, \Tr M_l \rho_l = 1$.
\item[(b)] $\rho \defeq \frac{1}{n} \cdot \sum_l \rho_l 
                    =   \frac{1}{m} \cdot I$,
           where $I$ is the identity operator on $\C^m$.
\item[(c)] $\forall l \, S(\rho_l \| \rho) = k$.
\item[(d)] For all $d$-dimensional subspaces $W$ of $\C^m$, 
           for all ordered sets of density matrices 
	   $\{\sigma_l\}_{l \in [n]}$ with support in $W$,
           $|\{l: \Tr M_l \sigma_l \leq 1/10\}| \geq n/4$.
\end{enumerate}
\end{result}
\paragraph{Remark: } The above result intuitively says that
the states $\rho_l$ on $\log m$ qubits cannot be compressed to less
than $\log d$ qubits with respect to the measurements $M_l$.

\subsection {Organisation of the rest of the paper } 
Section~\ref{sec:prelim} defines several basic concepts which
will be required for the proofs of the main results. 
In Section~\ref{sec:simul}, we prove a version of the message
compression result
for bounded error private coin simultaneous message protocols and
state the direct sum result for such protocols.  Our version is
slightly stronger than the one in~\cite{chakra:simul}.  The main ideas
of this work (i.e.{} the use of the Substate theorem and rejection
sampling) are already encountered in this section. In
Section~\ref{sec:multirounds}, we prove the compression result for
$k$-round bounded error private coin protocols, and state the direct
sum result for such protocols. We prove 
the impossibility of quantum compression in 
Section~\ref{sec:quantumimposs}. Finally, we conclude by mentioning
some open problems in Section~\ref{sec:conclusion}.

\section{Preliminaries}
\label{sec:prelim}
\subsection{Information theoretic background}
In this paper, $\ln$ denotes the natural logarithm and $\log$ denotes
logarithm to base $2$. All random variables will have finite range.
Let $[k] \defeq \{1, \ldots, k\}$. Let 
$P, Q: [k] \rightarrow {\mathbb R}$.
The {\em total variation distance} (also known as
{\em $\ell_1$-distance}) between $P, Q$ is defined as 
$\totvar{P - Q} \defeq \sum_{i \in [k]} |P(i) - Q(i)|$. We
say $P \leq Q$ iff $P(i) \leq Q(i)$ for all $i \in [k]$.
Suppose $X, Y, Z$
are random variables with some joint distribution.
The {\em Shannon entropy} of $X$ is defined as 
$H(X) \defeq -\sum_{x} \Pr[X = x] \log \Pr[X = x]$. 
The mutual information of $X$ and $Y$ is
defined as $I(X:Y) \defeq H(X) + H(Y) - H(XY)$. For 
$z \in \range(Z)$, $I((X:Y) \mid Z=z)$ denotes the mutual information
of $X$ and $Y$ conditioned on the event $Z = z$ i.e. the mutual
information arising from the joint distribution of $X, Y$ conditioned
on $Z = z$. Define $I((X:Y) \mid Z) \defeq \E_{Z} I((X:Y) \mid Z=z)$.
It is readily seen that 
$I((X:Y) \mid Z) = H(XZ) + H(YZ) - H(XYZ) - H(Z)$.
For a good introduction to information theory, see
e.g.~\cite{cover:infotheory}.

We now recall the definition of
an important information theoretic quantity called
{\em relative entropy}, also known as {\em Kullback-Leibler
divergence}.
\begin{definition}[Relative entropy]
Let $P$ and $Q$ be probability distributions on a 
set $[k]$. The relative entropy of $P$ and $Q$ is given by
$ \displaystyle
S(P \| Q) \defeq \sum_{i\in [k]} P(i) \log \frac{P(i)}{Q(i)}$.
\end{definition}

The following facts follow easily from the definitions.
\begin{fact}
\label{fact:additivity}
Let $X, Y, Z, W$  be random variables with some joint distribution. 
Then, 
\begin{enumerate}
\item[(a)] $I(X : YZ) = I(X : Y) + I((X : Z) \mid Y)$;
\item[(b)] $I(XY : Z \mid W) \geq I(XY : Z) - H(W)$.
\end{enumerate}
\end{fact}

\begin{fact}
\label{fact:exprel}
Let $(X, M)$  be a pair of random variables with some joint
distribution. 
Let $P$ be the (marginal) probability distribution of $M$, and
for each $x \in \range(X)$, let $P_x$ be the conditional distribution
of $M$ given $X=x$. Then $I(X:M) = \E_X [S(P_x \| P)]$, 
where the expectation is taken according
to the marginal distribution of $X$.
\end{fact}
Thus, if $I(X:M)$ is small, then we can conclude that 
$S(P_x \| P)$ is small on the average. 
  
Using Jensen's inequality, one can derive the following property of
relative entropy.
\begin{fact}[Monotonicity] 
\label{fact:monotonicity}
Let $P$ and $Q$ be probability
distributions on the set $[k]$ and $\event \subseteq [k]$. Let $D_P =
(P(\event), 1-P(\event))$ and $D_Q = (Q(\event), 1-Q(\event))$ be the
two-point distributions determined by $\event$. Then, 
$S(D_P \| D_Q) \leq S(P \| Q)$.  
\end{fact}

Our main information theoretic tool in this paper is the
following theorem (see \cite{jain:substate}). 
\begin{fact}[Substate theorem]
\label{fact:substate}
Suppose $P$ and $Q$ are probability distributions on $[k]$ such that
$S(P \| Q) = a$. Let $r \geq 1$. Then,
\begin{enumerate}
\item[(a)] the set $\Good \defeq \{i \in [k]:
\frac{P(i)}{2^{r(a+1)}} \leq Q(i)\}$ has probability at least
$1-\frac{1}{r}$ in $P$;
\item[(b)] There is a distribution $\widetilde{P}$ on $[k]$ 
such that $\totvar{P-\widetilde{P}} \leq \frac{2}{r}$ and
$\alpha \widetilde{P} \leq Q$, where
$\alpha \defeq \left(\frac{r-1}{r}\right)2^{-r(a+1)}$.
\end{enumerate}
\end{fact}
\begin{proof}
Let $\Bad\defeq [k]-\Good$. 
Consider the two-point distributions $D_P=(P(\Good), 1-P(\Good))$ and
$D_Q=(Q(\Good), 1-Q(\Good))$. By Fact~\ref{fact:monotonicity}, 
$S(D_P \| D_Q) \leq a$, that is,
\[ 
P(\Good) \log \frac{P(\Good)}{Q(\Good)} + P(\Bad) \log
\frac{P(\Bad)}{Q(\Bad)} \leq a.
\]
From our definition, $P(\Bad)/Q(\Bad) > 2^{r(a+1)}$.
Now, $P(\Good) \log \frac{P(\Good)}{Q(Good)} \geq P(\Good)
\log P(\Good) > -1$ (because $x\log x \geq (- \log e)/e > -1$ for $0
\leq x \leq 1$). It follows that $P(\Bad) \leq \frac{1}{r}$, thus
proving part (a).
Let $\widetilde{P}(i) \defeq P(i)/P(\Good)$ for $i \in \Good$ and
$\widetilde{P}(i)=0$ otherwise. Then, $\widetilde{P}$ satisfies the
requirements for part (b).
\end{proof}

\subsection {Chernoff-Hoeffding bounds} We will need the following
standard Chernoff-Hoeffding 
bounds on tails of probability distributions of sequences of
bounded, independent, identically distributed random variables.
Below, the notation $B(t,q)$ stands for the binomial distribution
got by $t$ independent coin tosses of a binary coin with success
probability $q$ for each toss. 
A {\em randomised predicate} $S$ on $[k]$ is a function 
$S: [k] \rightarrow [0, 1]$. 
For proofs of the following bounds,
see e.g. \cite[Corollary A.7, Theorem A.13]{AlSp}.

\begin{fact}
\label{fact:chernoff}
\mbox{}
\begin{enumerate}
\item [(a)] Let $P$ be a probability
distribution on $[k]$ and $S$ a randomised
predicate on $[k]$. Let 
$\displaystyle p \defeq \E_{x \in_P [k]} [S(x)]$. 
Let ${\mathbf Y} \defeq \tuple{Y_1, \ldots, Y_r}$ be a sequence of $r$
independent random variables, each with distribution $P$. Then,
\[ 
\Pr_{\bY} [|\E_{i \in_U [r]}[ S(Y_i)] - p| > \epsilon] 
         < 2\exp(-2\epsilon^2 r).
\]
\item[(b)] Let $R$ be a random variable with binomial distribution
$B(t,q)$. Then, 
\[ 
\Pr[R < \frac{1}{2} tq] < \exp\left(-\frac{1}{8} tq\right). 
\]
\end{enumerate}
\end{fact}

\subsection{Communication complexity background}
In the two-party private
coin randomised communication complexity model~\cite{yao:cc}, two 
players
Alice and Bob are required to collaborate to compute a function $f :
{\cal X} \times {\cal Y} \rightarrow \cZ$. Alice is given $x \in
\cX$ and Bob is given $y \in \cY$. Let $\Pi(x,y)$ be the random
variable denoting the entire 
transcript of the messages exchanged by Alice
and Bob by following the protocol $\Pi$ on input $x$ and $y$. We say
$\Pi$ is a $\delta$-error protocol if for all $x$ and $y$, the answer
determined by  the players is correct with probability (taken over the
coin tosses of Alice and Bob) at least $1-\delta$. The communication
cost of $\Pi$ is the maximum length of $\Pi(x,y)$
over all $x$ and $y$, and over all random choices of Alice and
Bob. The $k$-round $\delta$-error private coin randomised communication 
complexity of $f$,
denoted $R^k_\delta(f)$, is the communication cost of the best
private coin $k$-round
$\delta$-error protocol for $f$. When $\delta$ is omitted, we mean that
$\delta=\frac{1}{3}$. 

We also consider private coin randomised simultaneous protocols
in this paper.
$R^{{\rm sim}}_\delta(f)$ denotes 
the $\delta$-error private coin randomised simultaneous communication 
complexity of $f$.
When $\delta$ is omitted, we mean that
$\delta=\frac{1}{3}$. 

Let $\mu$ be a probability distribution on $\cX \times \cY$. 
A deterministic protocol $\Pi$ has distributional 
error $\delta$ if the probability of correctness
of $\Pi$, averaged with respect to $\mu$, is least $1 - \delta$.
The $k$-round $\delta$-error distributional communication 
complexity of $f$, denoted $C^k_{\mu, \delta}(f)$, 
is the communication cost of the best $k$-round deterministic
protocol for $f$ with distributional error $\delta$. 
$\mu$ is said to be a product distribution if there exist
probability distributions $\mu_{\cX}$ on $\cX$ and $\mu_{\cY}$
on $\cY$ such that $\mu(x, y) =\mu_{\cX}(x) \cdot \mu_{\cY}(y)$
for all $(x, y) \in \cX \times \cY$.
The $k$-round $\delta$-error
product distributional communication complexity of $f$ is
defined as $C^k_{[\,], \delta}(f) = \sup_{\mu} C^k_{\mu, \delta}(f)$, 
where the supremum is taken over all product distributions $\mu$ on 
$\cX \times \cY$. When $\delta$ is omitted, we mean that
$\delta=\frac{1}{3}$. 

We now recall the definition of the important notion of 
{\em information cost} of a communication protocol from
Bar-Yossef et al.~\cite{bar:ic}.

\begin{definition}[Information cost] Let $\Pi$ be a
private coin randomised protocol for a function 
$f: {\cal X} \times {\cal Y}
\rightarrow {\cal Z}$. Let $\Pi(x,y)$ be the entire message
transcript of the protocol on
input $(x,y)$. Let $\mu$ be a distribution on ${\cal X}
\times {\cal Y}$, and let the input random variable
$(X, Y)$ have distribution $\mu$. The
{\em information cost of  $\Pi$ under $\mu$} is defined to be 
$I(XY : \Pi(X,Y))$. 
The $k$-round $\delta$-error information complexity of 
$f$ under the
distribution $\mu$, denoted by $\IC^k_{\mu, \delta}(f)$, is the infimum
information cost under $\mu$ of a $k$-round
$\delta$-error protocol for $f$.
$\IC^{{\rm sim}}_\delta(f)$ denotes the infimum information cost 
under the uniform probability distribution on the inputs
of a private coin simultaneous $\delta$-error protocol for $f$.
\end{definition}
\paragraph{Remark: }
In Chakrabarti et al.~\cite{chakra:simul}, the information cost
of a private coin $\delta$-error
simultaneous message protocol $\Pi$ is defined as follows: Let
$X$ ($Y$) denote the random variable corresponding to Alice's 
(Bob's) input, and let $M$ ($N$) denote the random variable 
corresponding to Alice's (Bob's) message to the referee. The
information cost of $\Pi$ is defined as I(X:M) + I(Y:N). We note 
that our definition of information cost coincides with 
Chakrabarti et al.'s definition for simultaneous message protocols.

Let $\mu$ be a probability distribution on $\cX \times \cY$.
The probability distribution $\mu^m$ on 
$\cX^m \times \cY^m$ is defined as 
$\mu^m(\tuple{x_1, \ldots, x_m}, \tuple{y_1, \ldots, y_m}) \defeq 
\mu(x_1, y_1) \cdot \mu(x_2, y_2) \cdots \mu(x_m, y_m)$. 
Suppose $\mu$ is a product probability distribution on 
$\cX \times \cY$.
It can be easily seen (see e.g.~\cite{bar:ic}) that for any 
positive integers 
$m, k$, and real $\delta > 0$, 
$IC^{k}_{\mu^m, \delta}(f^m) \geq m \cdot IC^{k}_{\mu, \delta}(f)$.
The reason for requiring $\mu$ to be a product distribution is as
follows. We define the notion of information cost for private coin
protocols only. This is because the proof of our message
compression theorem (Theorem~\ref{thm:multicompress}), which
makes use of information cost,
works for private coin protocols only.
If $\mu$ is not a product distribution, the
protocol for $f$ which arises out of the 
protocol for $f^m$ in the proof of the above inequality fails to
be a private coin protocol, even if the protocol for $f^m$ was
private coin to start with. 
To get over this restriction on $\mu$,
Bar-Yossef et al.~\cite{bar:ic} introduced the notion of
{\em conditional information cost} of a protocol.
Suppose the distribution $\mu$ is expressed as a convex combination
$\mu = \sum_{d \in K} \kappa_d \mu_d$ of product distributions
$\mu_d$, where $K$ is some finite index set.
Let $\kappa$ denote the probability distribution on $K$ defined
by the numbers $\kappa_d$.
Define the random variable $D$ to be distributed
according to $\kappa$. Conditioned on $D$, $\mu$ is a product
distribution on $\cX \times \cY$. We will call $\mu$ a mixture
of product distributions $\{\mu_d\}_{d \in K}$ and say that
$\kappa$ {\em partitions} $\mu$. 
The probability distribution $\kappa^m$ on 
$K^m$ is defined as 
$\kappa^m(d_1, \ldots, d_m) \defeq 
\kappa(d_1) \cdot \kappa(d_2) \cdots \kappa(d_m)$. 
Then $\kappa^m$ partitions $\mu^m$ in a natural way.
The random variable $D^m$ has distribution $\kappa^m$. Conditioned
on $D^m$, $\mu^m$ is a product distribution on
$\cX^m \times \cY^m$.
\begin{definition}[Conditional information cost] Let $\Pi$ be a
private coin randomised protocol for a function 
$f: {\cal X} \times {\cal Y}
\rightarrow {\cal Z}$. Let $\Pi(x,y)$ be the entire message
transcript of the protocol on
input $(x,y)$. Let $\mu$ be a distribution on ${\cal X}
\times {\cal Y}$, and let the input random variable
$(X, Y)$ have distribution $\mu$. Let $\mu$ be a mixture of
product distributions partitioned by $\kappa$. Let the random variable
$D$ be distributed according to $\kappa$. The
{\em conditional information cost of  $\Pi$ under $(\mu, \kappa)$} 
is defined to be $I((XY : \Pi(X,Y)) \mid D)$. 
The $k$-round $\delta$-error conditional information complexity of 
$f$ under $(\mu, \kappa)$, denoted by 
$\IC^k_{\mu, \delta}(f \mid \kappa)$, is the infimum
conditional information cost under $(\mu, \kappa)$ of a $k$-round
$\delta$-error protocol for $f$.
\end{definition}

The following
facts follow easily from the results in Bar-Yossef et al.~\cite{bar:ic}
and Fact~\ref{fact:additivity}.
\begin{fact}
Let $\mu$ be a probability distribution on $\cX \times \cY$.
Let $\kappa$ partition $\mu$.
For any $f: \cX \times \cY \rightarrow {\cal Z}$, positive integers 
$m, k$, real $\delta > 0$, 
$
IC^{k}_{\mu^m, \delta}(f^m \mid \kappa^m) \geq 
m \cdot IC^{k}_{\mu, \delta}(f \mid \kappa) \geq
m \cdot (IC^{k}_{\mu, \delta}(f) - H(\kappa)).
$
\label{fact:addinfo}
\end{fact}
\begin{fact} 
With the notation and assumptions of Fact~\ref{fact:addinfo},
$R^k_\delta (f) \geq IC^k_{\mu, \delta}(f \mid \kappa)$.
\label{fact:CgeqIC}
\end{fact}

\subsection{Sampling uniformly random orthonormal sets of vectors}
To prove our result about the incompressibility of quantum information,
we need to define
the notion of a uniformly random set of size $d$
of orthonormal vectors from $\C^m$. Let $\bU(m)$ denote the group
(under matrix multiplication) of $m \times m$ complex unitary matrices.
Being a compact topological group, it has a unique Haar probability
measure on its
Borel sets which is both left and right invariant under multiplication
by unitary matrices 
(see e.g.~[Chapter~14, Corollary~20]\cite{royden:analysis}).
Let $\bU_{m,d}$, ($1 \leq d \leq m$) denote the topological 
space of $m \times d$ complex matrices
with orthonormal columns. $\bU_{m,d}$ is compact, and
the group $\bU(m)$ acts on $\bU_{m,d}$ via multiplication from the left.
Let $f_{m, d}: \bU(m) \rightarrow \bU_{m,d}$ be the map 
got by discarding
the last $m - d$ columns of a unitary matrix. $f_{m, d}$ induces
a probability measure $\bmu_{m,d}$ on 
the Borel sets of $\bU_{m,d}$ from the Haar probability measure on
$\bU(m)$. $\bmu_{m,d}$ is invariant under the 
action of $\bU(m)$, and is in fact the unique $\bU(m)$-invariant
probability measure on the Borel sets of $\bU_{m,d}$
(see e.g.~[Chapter~14, Theorem~25]\cite{royden:analysis}).
By a uniformly random ordered set $(v_1, \ldots, v_d)$, 
$1 \leq d \leq m$ of orthonormal vectors from
$\C^m$, we mean an element of $\bU_{m,d}$ chosen according to 
$\bmu_{m,d}$.
By a uniformly random $d$ dimensional subspace $V$ of $\C^m$, we mean a
subspace  $V \defeq \linspan (v_1, \ldots, v_d)$, where
$(v_1, \ldots, v_d)$ is a uniformly random ordered set
of orthonormal vectors from $\C^m$.

Let $\bO(m)$ denote the group
(under matrix multiplication) of $m \times m$ real orthogonal matrices.
Identify $\C^m$ with $\R^{2m}$ by treating a complex number as a 
pair of real numbers.
A uniformly random unit vector in $\C^m$ (i.e. a vector distributed
according to $\bmu_{m,1}$) is the same as a uniformly random unit
vector in $\R^{2m}$, since
$\bU(m)$ is contained in $\bO(2m)$. From now on, while considering
metric and measure theoretic properties of $\bU_{m,1}$, it may help
to keep the above identification of $\C^m$ and $\R^{2m}$ in mind.

One way of generating a uniformly random unit vector in $\R^m$ is
as follows: First choose $\tuple{y_1, \ldots, y_m}$ independently, 
each $y_i$ being chosen according to the one dimensional
Gaussian distribution with mean $0$ and variance $1$ (i.e. a 
real valued random variable with probability density function 
$\frac{\exp(-y^2)}{\sqrt{2 \pi}}$). Normalise to get the unit
vector $\tuple{x_1, \ldots, x_m}$, where 
$x_i \defeq \frac{y_i}{\sqrt{y_1^2 + \cdots + y_m^2}}$ (note that
any $y_i = 0$ with zero probability). It is easily seen
that the resulting distribution on unit vectors is $\bO(m)$-invariant,
and hence, the above process generates a uniformly random unit vector
in $\R^m$.

From the above discussion, one can prove the following fact.
\begin{fact}
\label{fact:symmetry}
\mbox{}
\begin{enumerate}
\item[(a)]
Let $1 \leq d \leq m$. Let $(v_1, \ldots, v_d)$ be 
distributed according to $\bmu_{m,d}$.
Then for each $i$, $v_i$ is distributed according to $\bmu_{m,1}$,
and for each $i, j$, $i \neq j$, $(v_i, v_j)$ is distributed according
to $\bmu_{m,2}$, 
\item[(b)]
Suppose $x, y$ are independent unit vectors, each distributed
according to $\bmu_{m,1}$. Let
$w'' \defeq y - \braket{x}{y} x$, and set $w \defeq x$ and
$w' \defeq \frac{w''}{\norm{w''}}$ (note that
$w'' = 0$ with probability zero).
Then the pair $(w, w')$ is
distributed according to $\bmu_{m,2}$. 
\item[(c)]
Suppose $x, y$ are independent unit vectors, each distributed
according to $\bmu_{m,1}$.
Let $V$ be a subspace of
$\C^m$ and define $\widehat{x} \defeq \frac{Px}{\norm{Px}}$,
$\widehat{y} \defeq \frac{Py}{\norm{Py}}$, where $P$ is the orthogonal
projection operator onto $V$ (note that $Px = 0$, $Py = 0$ are
each zero probability events). Then $\widehat{x}, \widehat{y}$ are
uniformly random independent unit vectors in $V$.
\end{enumerate}
\end{fact}

We will need to 
`discretise' the set of $d$-dimensional subspaces of $\C^m$.
The discretisation is done by using a {\em $\delta$-dense} subset of 
$\bU_{m,1}$. A subset ${\cal N}$ of $\bU_{m,1}$
is said to be $\delta$-dense if each vector $v \in \bU_{m,1}$
has some vector in ${\cal N}$ at distance no larger than $\delta$ from
it. We require the following fact about $\delta$-dense subsets of
$\bU_{m,1}$.
\begin{fact}[\mbox{\cite[Lemma~13.1.1, Chapter~13]{matousek:dg}}]
\label{fact:net}
For each $0 < \delta \leq 1$, there is a $\delta$-dense subset
${\cal N}$ of $\bU_{m,1}$ satisfying 
$|{\cal N}| \leq (4/\delta)^{2m}$.
\end{fact}

A mapping $f$ between two metric spaces is said to be
{\em $1$-Lipschitz} if the distance between $f(x)$ and $f(y)$ is
never larger than the distance between $x$ and $y$. The following
fact says that a $1$-Lipschitz function 
$f: \bU_{m,1} \rightarrow \R$ greatly exceeds its expectation with
very low probability.
It follows by combining Theorem~14.3.2 and Proposition~14.3.3 of
\cite[Chapter 14]{matousek:dg}.
\begin{fact}
\label{fact:conc}
Let $f: \bU_{m,1} \rightarrow \R$ be $1$-Lipschitz. Then for all
$0 \leq t \leq 1$, 
$\Pr[f > \E[f] + t + 12/\sqrt{2m}] \leq 2 \exp(-t^2 m)$.
\end{fact}

\subsection{Quantum information theoretic background}
We consider a quantum system with Hilbert space $\C^m$. For $A, B$
Hermitian operators on $\C^m$, $A \leq B$ is a shorthand for
the statement ``$B - A$ is positive semidefinite''.
A {\em POVM element} $M$ over $\C^m$ is a Hermitian operator satisfying
the property $0 \leq M \leq I$, where $0, I$ are the zero and identity
operators respectively on $\C^m$.
For a POVM element $M$ over $\C^m$ and a subspace $W$ of $\C^m$, define
$\displaystyle
M(W) \defeq \max_{w \in W: \norm{w} = 1} \unibraket{w}{M}{w}$.
For subspaces $W, W'$ of $\C^m$, define 
$\Delta(W, W') \defeq \max_M |M(W) - M(W')|$, where the maximum
is taken over all POVM elements $M$ over $\C^m$. $\Delta(W, W')$
is a measure of how well one can distinguish between subspaces $W, W'$
via a measurement. For a good introduction to quantum information
theory, see \cite{nielsen:quant}.

The following fact can be proved from the results in
\cite{aharonov:mixed}.
\begin{fact}
\label{fact:povmtrdist}
Let $M$ be a POVM element over $\C^m$ and let
$w, \widehat{w} \in \C^m$ be unit vectors. Then,
$|\unibraket{w}{M}{w}-\unibraket{\widehat{w}}{M}{\widehat{w}}| 
\leq \norm{w-\widehat{w}}$. 
\end{fact}

A density matrix $\rho$ over $\C^m$ is a Hermitian, positive 
semidefinite operator on $\C^m$ with unit trace.
If $A$ is a quantum system with Hilbert space $\C^m$ having density 
matrix $\rho$, then
$S(A) \defeq S(\rho) \defeq -\Tr \rho \log \rho$ is the
{\em von Neumann entropy} of $A$.
If $A, B$ are two
disjoint quantum systems, the {\em mutual information}
of $A$ and $B$ is defined as $I(A : B) \defeq S(A) + S(B) - S(AB)$.
For density matrices $\rho, \sigma$ over $\C^m$, their 
{\em relative entropy} is defined as 
$S(\rho \| \sigma) \defeq \Tr \rho(\log \rho - \log \sigma)$.
Let $X$ be a classical random variable with finite range and
$M$ be a $m$-dimensional quantum encoding of $X$ i.e. for every
$x \in \range(X)$ there is a density matrix $\sigma_x$ over $\C^m$ 
($\sigma_x$ represents a `quantum encoding' of $x$). 
Let $\sigma \defeq \E_X \sigma_x$, where the expectation is taken
over the (marginal) probability distribution of $X$.
Then, $I(X:M) = \E_X S(\sigma_x \| \sigma)$.

\section{Simultaneous message protocols}
\label{sec:simul}
In this section, we prove a result of ~\cite{chakra:simul}, which
states that if the mutual information between the message and the
input is at most $k$, then the protocol can be modified so that the
players send messages of length at most $O(k + \log n)$ bits. Our
proof will make use of the Substate Theorem and a rejection sampling
argument. In the next section, we will show how to extend this
argument to multiple-round protocols.

Before we formally state the result and its proof, let us outline the
main idea. Fix a simultaneous message protocol for computing the
function $f: \{0, 1\}^n \times \{0, 1\}^n \rightarrow \cZ$.
Let $X \in_U \{0,1\}^n$. Suppose $I(X:M) \leq a$, where $M$ be the
message sent by Alice to the referee when her input is $X$.  Let
$s_{xy}(m)$ be conditional probability that the referee computes
$f(x,y)$ correctly when Alice's message is $m$, her input is $x$ 
and Bob's input is $y$. 

We want to show that we can choose a small subset $\cM$ of possible
messages, so that for most $x$, Alice can generate a message $M'_x$
from this subset (according to some distribution that depends on $x$),
and still ensure that $\E[s_{xy}(M'_x)]$ is close to 1, for all $y$.
Let $P_x$ be the distribution of $M$ conditioned on the event $X=x$.
For a fixed $x$, it is possible to argue that we can confine Alice's
messages to a certain small subset $\cM_x \subseteq [k]$.  Let $\cM_x$
consist of $O(n)$ messages picked according to the distribution
$P_x$. Then, instead of sending messages according to the distribution
$P_x$, Alice can send a random message chosen from $\cM_x$. Using
Chernoff-Hoeffding bounds one can easily verify that $\cM_x$ will
serve our purposes with exponentially high probability.

However, what we really require is a set of samples $\{\cM_x\}$ whose
union is small, so that she and the referee can settle on a
common succinct encoding for the messages.  Why should such samples
exist?  Since $I(X:M)$ is small, we have by Fact~\ref{fact:exprel}
that for most $x$, the relative entropy $S(P_x \| Q)$ is bounded (here
$Q$ is the distribution of the message $M$, i.e., $Q=
\E_X[P_X]$). By combining this fact, the Substate Theorem
(Fact~\ref{fact:substate}) and a {\em rejection sampling} argument
(see e.g.~\cite[Chapter 4, Section 4.4]{ross:simul}), one can show
that if we choose a sample of messages according to the distribution
$Q$, then, for most $x$, roughly one in every $2^{O(a)}$ messages `can
serve' as a message sampled according to the distribution $P_x$. 
Thus, if we
pick a sample of size $n \cdot 2^{O(a)}$ according to $Q$, then for
most $x$ we can get a the required sub-sample $\cM_x$.
of $O(n)$ elements. The formal arguments are presented below.

The following easy lemma is the basis of the rejection sampling
argument. 
\begin{lemma}[Rejection sampling]
\label{lm:onecopy}
Let $P$ and $Q$ be probability distributions on $[k]$ 
such that $2^{-a}P \leq Q$. Then, there exist correlated
random variables $X$ and
$\chi$ taking values in $[k] \times \{0,1\}$, such that: (a) $X$ has
distribution $Q$, (b) $\Pr[\chi=1]=2^{-a}$ and (c) $\Pr[X=i \mid
\chi=1] = P(i)$.
\end{lemma}
\begin{proof}
Since the distribution of $X$ is required to be $Q$, we will just
describe the conditional distribution of $\chi$ for each 
potential value $i$ for $X$: let $\Pr[\chi=1 \mid X=i] =
P(i)/(2^aQ(i))$. Then, 
\[
\Pr[\chi=1] = \sum_{i \in [k]} P[X=i] \cdot \Pr[\chi=1 \mid X=i] =
2^{-a}
\] 
and
\[
\Pr[ X=i \mid \chi=1] = \frac{\Pr[X=i \wedge \chi=1]}{\Pr[\chi=1]} =
\frac{Q(i) \cdot P(i)/(2^aQ(i))}{2^{-a}} = P(i).
\]
\end{proof}

In order to combine this argument with the Substate Theorem to
generate simultaneously a sample $\cM$ of messages according to the
distribution $Q$ and several subsamples $\cM_x$, we will need a slight
extension of the above lemma. 
\begin{lemma} 
\label{lem:correlatedsampling}
Let $P$ and $Q$ be probability distributions on $[k]$ such that
$2^{-a}P \leq Q$. Then, for each integer $t\geq 1$, there exist 
correlated random variables ${\mathbf X}=\tuple{X_1,X_2,\ldots,X_t}$
and ${\mathbf Y}=\tuple{Y_1,Y_2,\ldots,Y_R}$ such that
\begin{enumerate}
\item[(a)] The random variables $(X_i: i\in [t])$ 
are independent and each $X_i$ has distribution $Q$; 
\item[(b)] $R$ is a random variable with binomial distribution
$B(t,2^{-a})$;  
\item[(c)] Conditioned on the event $R=r$, the random variables $(Y_i: i
\in [r])$ are independent and each $Y_i$ has distribution $P$.
\item[(d)] ${\mathbf Y}$ is a subsequence of ${\mathbf X}$ (with
probability 1).
\end{enumerate}
\end{lemma}
\begin{proof}
We generate $t$ independent copies of the random variables $(X,\chi)$
promised by Lemma~\ref{lm:onecopy}; this gives us ${\mathbf
X}=\tuple{X_1,X_2,\ldots,X_t}$ and ${\mathbf
\chi}=\tuple{\chi_1,\chi_2,\ldots,\chi_t}$. Let ${\mathbf Y}
\defeq \tuple{X_i : \chi_i=1}$. It is easy to verify that ${\mathbf
X}$ and ${\mathbf Y}$ satisfy conditions (a)--(d).
\end{proof}

Our next lemma uses Lemma~\ref{lem:correlatedsampling} to pick a
sample of messages according to the average distributions $Q$ and
find sub-samples inside it for several distributions $P_x$. This lemma
will be crucial to show the compression result for simultaneous
message protocols (Theorem~\ref{thm:simulcompress}). 
\begin{lemma} 
\label{lem:sampling1}
Let $Q$ and $P_1,P_2,\ldots,P_N$ be probability distributions on
$[k]$. Define $a_i \defeq S(P_i \| Q)$.  Suppose $a_i < \infty$ for
all $i \in [N]$.  Let $s_{ij},s_{ij},\ldots,s_{ij}$ be functions from
$[k]$ to $[0,1]$. (In our application, they will correspond 
to conditional
probability that the referee gives the correct answer when Alice sends
a certain message from [k]).  Let $\displaystyle p_{ij} \defeq E_{y
\in_{P_i}[k]} [ s_{ij}(y)]$.  Fix $\epsilon \in (0,1]$.  Then, there
exists a sequence $\bx \defeq \tuple{x_1, \ldots, x_t}$ of elements of
$[k]$ and subsequences $\by^1, \ldots, \by^N$ of $\bx$ such that
\begin{enumerate}
\item[(a)] $\by^i$ is a subsequence of
$\tuple{x_1, \ldots, x_{t_i}}$ where, 
\( \textstyle ~
t_i \defeq \left\lceil\frac{8 \cdot 2^{(a_i+1)/\epsilon} 
                               \cdot \log(2N)}  
                            {(1 - \epsilon) \epsilon^2} \right\rceil.
\)
\item[(b)] For $i,j=1,2,\ldots,N$, 
$\displaystyle 
 \left|\E_{\ell \in_U [r_i]} [s_{ij}(\by^i[\ell])] - p_{ij}\right| 
             \leq 2 \epsilon$, 
where $r_i$ is the length of $\by^i$.
\item[(c)] $t \defeq \max_i t_i$.
\end{enumerate}
\end{lemma}
\begin{proof}
Using part (b) of Fact~\ref{fact:substate}, 
we obtain  distributions $\widetilde{P}_i$ such that 
\[
\forall i \in [k], ~ \totvar{P_i - \widetilde{P}_i} \leq 2 \epsilon 
~~~ {\rm and} ~~~
(1-\epsilon)2^{-(a_i+1)/\epsilon}\widetilde{P}_i \leq Q. 
\]
Using Lemma~\ref{lem:correlatedsampling}, 
we can construct correlated
random variables $(\bX,\bY^1,\bY^2,\ldots,\bY^N)$ such that $\bX$ is
a sequence of $t \defeq \max_i t_i$ independent random variables, each
distributed according to $Q$, and $(\bX[1,t_i],\bY^i)$ satisfying
conditions (a)--(d) (with $P=P_i$, $a=(a_i+1)/\epsilon - \log
(1-\epsilon)$ and 
$t=t_i$). We will show that with non-zero probability these random
variables satisfy conditions (a) and (b) of the present
lemma. This implies
that there is a choice $(\bx, \by^1, \ldots, \by^N)$ for 
$(\bX, \bY^1, \ldots, \bY^N)$ satisfying
parts (a) and (b) of the present lemma.

Let $R_i$ denote the length of $\bY^i$. 
Using part~(b) of Fact~\ref{fact:chernoff}, 
$\Pr[\exists i, R_i < (4/\epsilon^2) \log(2 N)] < N
\cdot \frac{1}{2N} =
\frac{1}{2}$. 
Now, condition on the event $R_i \geq
\left(\frac{4}{\epsilon^2}\right) \log (2N)$, for all 
$1 \leq i \leq N$. 
Define $\displaystyle \widetilde{p}_{ij} \defeq 
\Pr_{y \in_{\widetilde{P}_i}[k]} [ s_{ij}(y)]$.
We use part~(a) of Fact~\ref{fact:chernoff} to 
conclude that 
\begin{equation}
\Pr_{\bY^i} \left[\left|\E_{\ell \in_U [r_i]}[s_{ij}(\bY^i[\ell])] 
              - \widetilde{p}_{ij}\right| > \epsilon\right] 
   < \frac{2}{(2N)^8}, ~~~ \forall i,j = 1, \ldots, N,
\label{firsttail}
\end{equation}
implying that
\begin{equation}
\Pr_{\bY^1, \ldots, \bY^N} 
    \left[\exists i,j, \, \left|\E_{\ell \in_U [r_i]} [s_{ij}(\bY^i[l])] 
		      - \widetilde{p}_{ij}\right| > \epsilon\right] 
\leq N^2 \times \frac{2}{(2N)^8}  < \frac{1}{2}. 
\label{secondtail}
\end{equation}
From (\ref{firsttail}), (\ref{secondtail}) and 
the fact that 
$\forall i,j ~~ |p_{ij} - \widetilde{p}_{ij}| \leq \epsilon$ 
(since $\totvar{P_i - \widetilde{P}_i} \leq 2\epsilon$), it
follows that part $(b)$ of our lemma holds with non-zero
probability. Part $(a)$ is never violated. 
Part $(c)$ is true by definition of $t$.  
\end{proof}

\begin{theorem}[Compression result, simultaneous messages] 
Suppose that
$\Pi$ is a $\delta$-error private coin simultaneous message
protocol for $f: \{0,1\}^n \times \{0,1\}^n \rightarrow \cZ$.
Let the inputs to $f$ be
chosen according to the uniform distribution. Let $X, Y$ denote
the random variables corresponding to Alice's and Bob's inputs
respectively, and $M_A, M_B$ denote the random variables corresponding
to Alice's and Bob's messages respectively. Suppose 
$I(X: M_A) \leq a$ and $I(Y:M_B) \leq b$. 
Then, there exist sets $\Good_A, \Good_B \subseteq \{0,1\}^n$ such 
that $|\Good_A| \geq \frac{2}{3} \cdot 2^n$ and  
$|\Good_B| \geq \frac{2}{3} \cdot 2^n $, and 
a private coin simultaneous message protocol $\Pi'$ 
with the following properties:
\begin{enumerate}
\item[(a)] In $\Pi'$, Alice sends messages of length at most 
$\frac{3a+1}{\epsilon} + \log (n+1) + 
 \log \frac{1}{\epsilon^2(1- \epsilon)} + 4$ 
bits and Bob sends messages of length at most 
$\frac{3b+1}{\epsilon} + \log (n+1) + 
 \log \frac{1}{\epsilon^2(1- \epsilon)} + 4$ 
bits.
\item[(b)] For each input $(x, y) \in \Good_A \times \Good_B$,
the error probability of $\Pi'$ is
at most $\delta + 4\epsilon$. 
\end{enumerate}
\label{thm:simulcompress}
\end{theorem}
\begin{proof}
Let $P$ be the distribution of $M_A$, and let
$P_x$ be its distribution under the condition $X=x$.  Note that by
Fact~\ref{fact:exprel}, we have $\E_X [S(P_x \| P)]\leq a$,
where the expectation is got by choosing $x$ uniformly from
$\{0, 1\}^n$.
Therefore there 
exists a set $\Good_A$, $|\Good_A| \geq \frac{2}{3} \cdot2^n$,
such that for all $x \in \Good_A, S(P_x \| P) \leq 3a$. 

Define 
$t_a \defeq \frac{8(n+1)2^{(3a+1)/\epsilon}}{\epsilon^2(1-\epsilon)}$.
From Lemma~\ref{lem:sampling1}, we know that  
there is a sequence of 
messages $\sigma = \tuple{m_1,\ldots, m_{t_a}}$
and subsequences $\sigma_x$ of
$\sigma$ such that on input $x \in \Good_A$, if Alice sends 
a uniformly chosen
random message of $\sigma_x$ instead of sending messages according to
distribution $P_x$, the probability of error for any $y \in \{0, 1\}^n$
changes by at most $2\epsilon$. 
We now define an intermediate protocol $\Pi''$ as follows.
The messages in $\sigma$ are encoded using at most
$\log t_a + 1$ bits. 
In protocol $\Pi''$ for $x \in \Good_A$, 
Alice sends a uniformly chosen random message from $\sigma_x$;
for $x \notin \Good_A$, Alice sends a fixed arbitrary message from
$\sigma$. Bob's strategy in $\Pi''$ is the same as in $\Pi$.
In $\Pi''$, the error probability of an input 
$(x, y) \in \Good_A \times \{0, 1\}^n$ is at most $\delta + 2\epsilon$,
and $I(Y: M_B) \leq b$.
Now arguing similarly,
the protocol $\Pi''$ can be converted to a protocol $\Pi'$ 
by compressing 
Bob's message to at most $\log t_b + 1$ bits, where 
$t_b \defeq \frac{8(n+1)2^{(3b+1)/\epsilon}}{\epsilon^2(1-\epsilon)}$.
In $\Pi'$, the error for an input 
$(x,y) \in \Good_A \times \Good_B$ is at most $\delta + 4\epsilon$.  
\end{proof}

\begin{corollary} 
Let $\delta, \epsilon > 0$.
Let $f: \{0,1\}^n \times \{0,1\}^n \rightarrow \cZ$ be a function.
Let the inputs to $f$ be
chosen according to the uniform distribution. 
Then there exist sets $\Good_A, \Good_B \subseteq \{0,1\}^n$ such 
that $|\Good_A| \geq \frac{2}{3} \cdot 2^n$, 
$|\Good_B| \geq \frac{2}{3} \cdot 2^n $, and 
\( \textstyle 
IC^{{\rm sim}}_{\delta} (f) \geq
\frac{\epsilon}{3}(R^{{\rm sim}}_{\delta + 4\epsilon} (f') - 
2 \log (n+1) - 2 \log \frac{1}{\epsilon^2(1- \epsilon) } - 
\frac{2}{\epsilon} - 8),
\) 
where $f'$ is the restriction of $f$ to $\Good_A \times \Good_B$.
\label{cor:ICgeqCSimul}
\end{corollary}

We can now prove the key theorem of Chakrabarti 
et al.~\cite{chakra:simul}.
\begin{theorem}[Direct sum, simultaneous messages] 
Let $\delta, \epsilon > 0$.
Let $f: \{0,1\}^n \times \{0,1\}^n \rightarrow \cZ$ be a function.
Define $\tilde{R}^{{\rm sim}}_{\delta} (f) \defeq 
\min_{f'} R^{{\rm sim}}_{\delta} (f')$, where the minimum is
taken over all functions $f'$ which are the restrictions of $f$
to sets of the form $A \times B$, $A, B \subseteq \{0, 1\}^n$, 
$|A| \geq \frac{2}{3} \cdot 2^n$, $|B| \geq \frac{2}{3} \cdot 2^n$.
Then,
\( \textstyle
R^{{\rm sim}}_{\delta}(f^m) \geq 
\frac{m\epsilon}{3}(\tilde{R}^{{\rm sim}}_{\delta + 4\epsilon} (f) - 
2 \log (n+1) - 2 \log \frac{1}{\epsilon^2(1- \epsilon) } - 
\frac{2}{\epsilon} - 8).
\) 
\label{thm:directsumsimul}
\end{theorem}
\begin{proof} 
Immediate from Fact~\ref{fact:CgeqIC}, Fact~\ref{fact:addinfo} and  
Corollary~\ref{cor:ICgeqCSimul}.
\end{proof} 
\paragraph{Remarks:} 
\ \\
1. \ The above theorem implies 
lower bounds for the simultaneous direct sum complexity of equality,
as well as lower bounds for some related problems as in
Chakrabarti et al.~\cite{chakra:simul}. The dependence of the bounds
on $\epsilon$ is better in our version. \\
2. \ A very similar direct sum theorem can be proved about 
two-party one-round private coin protocols. \\
3. \ All the results in this section, including the above remark,
hold even when $f$ is a relation.

\section {Two-party multiple-round protocols}
\label{sec:multirounds}
We first prove Lemma~\ref{lem:correlatedlasvegas}, which intuitively
shows that if $P, Q$ are probability distributions on $[k]$ such that
$P \leq 2^aQ$, then about it is enough to sample $Q$ independently
$2^{O(a)}$ times to produce one sample element $Y$ according to
$P$. In the statement of the lemma, the random variable $\bX$
represents an infinite sequence of independent sample elements chosen
according to $Q$, the random variable $R$ indicates how many of these
elements have to be considered till `stopping'. $R = \infty$ indicates
that we do not `stop'. If we do `stop', then either we succeed in
producing a sample according $P$ (in this case, the sample $Y=X_R$),
or we give up (in this case, we set $Y=0$).  In the proof of the
lemma, $\star$ indicates that we do not `stop' at the current
iteration and hence the rejection sampling process must go further.
\begin{lemma} 
Let $P$ and $Q$ be probability distributions on $[k]$,
such that $\Good \defeq \{i \in [k]:
\frac{P(i)}{2^{a}} \leq Q(i)\}$ has probability exactly
$1-\epsilon$ in $P$. Then,  there exist correlated random variables 
$\bX \defeq \tuple{X_i}_{i \in \nat}$, $R$ and $Y$ such that
\begin{enumerate}
\item[(a)] the random variables $(X_i: i \in \nat)$ are independent and
           each has distribution $Q$;
\item[(b)] $R$ takes values in $\nat \cup \{\infty\}$ and
           $\E[R]=2^a$;
\item[(c)] if $R \neq \infty$, then $Y=X_R$ or $Y=0$;
\item[(d)] $Y$ takes values in $\{0\} \cup [k]$, such that: 
\(        \Pr[Y=i] = \left\{ 
		      \begin{array}{cl}
			P(i) & \mbox{if $i\in \Good$}\\
			0    & \mbox{if $i \in [k]-\Good$}\\
			\epsilon & \mbox{if $i=0$}.
		      \end{array}
		      \right.
\)
\end{enumerate}
\label{lem:correlatedlasvegas}
\end{lemma}
\begin{proof}
First, we define a pair of correlated random variables
$(X,Z)$, where $X$ takes values in $[k]$ and $Z$ in $[k]\cup
\{0,\star\}$. Let $P':[k] \rightarrow [0,1]$ be defined by
$P'(i)=P(i)$ for $i\in \Good$, and $P'(i)=0$ for $i \in [k]-\Good$.
Let $\beta \defeq \epsilon 2^{-a}/(1-(1-\epsilon)2^{-a})$ and
$\gamma_i\defeq P'(i)2^{-a}/Q(i)$. The joint probability distribution
of $X$ and $Z$ is given by
\[
\forall i \in [k], ~ \Pr[X=i] = Q(i) 
~~~ {\rm and} ~~~
\Pr[Z=j \mid X=i] = \left\{
\begin{array}{cl}
\gamma_i                     & {\rm if} j = i         \\
\beta(1-\gamma_i)            & {\rm if} j = 0         \\
1-\gamma_i-\beta(1-\gamma_i) & {\rm if} j = \star     \\
0                            & {\rm otherwise}.
\end{array}
\right.
\]
Note that this implies that 
\[
\Pr[Z \neq \star] = \sum_{i \in [k ]} Q(i)\cdot[\gamma_i+
\beta(1-\gamma_i)] = \beta + (1-\beta)\sum_{i \in [k]} P'(i)2^{-a}  
			   = \beta + (1-\beta) (1-\epsilon)2^{-a} =
2^{-a}. 
\]
Now, consider the sequence of random variables 
$\bX \defeq \tuple{X_i}_{i \in \nat}$ and 
$\bZ \defeq \tuple{Z_i}_{i \in \nat}$, where each 
$(X_i,Z_i)$ has the same
distribution as $(X,Z)$ defined above and $(X_i,Z_i)$ is independent
of all $(X_j,Z_j), j \neq i$. Let 
$R \defeq \min \{i: Z_i \neq \star\}$; 
$R \defeq \infty$ if $\{i: Z_i \neq \star\}$ is the empty set.
$R$ is a geometric random variable with success probability $2^{-a}$,
and so satisfies part~(b) of the present lemma.
Let $Y \defeq Z_R$ if $R \neq \infty$ and $Y \defeq 0$ if
$R = \infty$. Parts~(a) and (c) are satisfied by construction.

We now verify that part~(d) is satisfied. Since $\Pr[R = \infty] = 0$,
we see that
\begin{eqnarray*}
\Pr[Y=i] &=& \sum_{r \in \nat} \Pr[R=r] \cdot \Pr[Z_r=i \mid R=r] \\
         &=& \sum_{r \in \nat} \Pr[R=r] \cdot 
			       \Pr[Z_r=i \mid Z_r \neq \star] \\
         &=& \sum_{r \in \nat} \Pr[R=r] \cdot 
			       \frac{\Pr[Z_r=i]}{\Pr[Z_r \neq \star]}, 
\end{eqnarray*}
where the second equality follows from the independence of 
$(X_r, Z_r)$ from all $(X_j, Z_j), j \neq r$. 
If $i \in [k]$, we see that
\begin{eqnarray*}
\Pr[Y=i] &=& \sum_{r \in \nat} \Pr[R=r] \cdot 
			       \frac{\Pr[Z_r=i]}{\Pr[Z_r \neq \star]}\\
         &=& \sum_{r \in \nat} \Pr[R=r] \cdot 
			       \frac{\Pr[X_r=i] \cdot 
			       \Pr[Z_r=i \mid X_r=i]}
			       {\Pr[Z_r \neq \star]} \\
         &=& \sum_{r \in \nat} \Pr[R=r] \cdot 
			       \frac{Q(i) \gamma_i}{2^{-a}} \\
         &=& \sum_{r \in \nat} \Pr[R=r] P'(i) = P'(i).
\end{eqnarray*}
Thus, for $i \in \Good$, $\Pr[Y = i] = P(i)$, and for
$i \in [k] - \Good$, $\Pr[Y = i] = 0$.
Finally,
\begin{eqnarray*}
\Pr[Y=0] &=& \sum_{r \in \nat} \Pr[R=r] \cdot 
			       \frac{\Pr[Z_r=0]}{\Pr[Z_r \neq \star]}\\
         &=& \sum_{r \in \nat} \frac{\Pr[R=r]}{2^{-a}}
             \sum_{j \in [k]}  \Pr[X_r=j] \cdot 
			       \Pr[Z_r=0 \mid X_r=j] \\
         &=& \sum_{r \in \nat} \frac{\Pr[R=r]}{2^{-a}}
             \sum_{j \in [k]}  Q(j) \cdot \beta (1 - \gamma_j) \\
         &=& \sum_{r \in \nat} \Pr[R=r] \epsilon = \epsilon.
\end{eqnarray*}
\end{proof}

Lemma~\ref{lem:sampling2} follows from
Lemma~\ref{lem:correlatedlasvegas}, and 
will be used to prove the message
compression result for two-party multiple-round protocols
(Theorem~\ref{thm:multicompress}).
\begin{lemma}
Let $Q$ and $P_1, \ldots, P_N$ be probability
distributions on $[k]$. Define $S(P_i \| Q)=a_i$. 
Suppose $a_i < \infty$ for all $i \in [N]$.
Fix $\epsilon \in (0,1]$.
Then, there exist random variables $\bX = \tuple{X_i}_{i \in \nat}$,
$R_1, \ldots, R_N$ and $Y_1, \ldots, Y_N$ such that
\begin{enumerate}
\item[(a)] $(X_i: i \in \nat)$ are independent random variables, each
           having distribution $Q$;
\item[(b)] $R_i$ takes values in $\nat \cup \{\infty\}$ and
           $\E[R_i]=2^{(a_i+1)/\epsilon}$; 
\item[(c)] $Y_j$ takes values in $[k] \cup \{0\}$, and
           there is a set $\Good_j \subseteq [k]$ with 
	   $P_j(\Good_j) \geq 1-\epsilon$ such that for all 
	   $\ell \in \Good_j$, $\Pr[Y_j=\ell] = P_j(\ell)$, for 
	   all $\ell \in [k]-\Good_j$, $\Pr[Y_j=\ell]=0$ and 
	   $\Pr[Y_j = 0] = 1- P_j(\Good_j) \leq \epsilon$; 
\item[(d)] if $R_j < \infty$, then $Y_j=X_{R_j}$ or $Y = 0$. 
\end{enumerate}
\label{lem:sampling2}
\end{lemma}
\begin{proof}
Using part~(a) of Fact~\ref{fact:substate}, 
we obtain for $j = 1, \ldots, N$, a set
$\Good_j \subseteq [k]$ such that $P_j(\Good_j) \geq 1-\epsilon$ and
$P_j(i) 2^{-(a_j+1)/\epsilon} \leq Q(i)$ for all $i \in \Good_j$. 
Now from Lemma~\ref{lem:correlatedlasvegas}, we can construct
correlated random variables $\bX$, $Y_1, \ldots, Y_N$, and
$R_1, \ldots, R_N$  satisfying the requirements of the present lemma.
\end{proof}

\begin{theorem}[Compression result, multiple rounds]
Suppose $\Pi$ is a $k$-round private coin randomised protocol for 
$f: \cX \times \cY \rightarrow \cZ$. Let the average
error of $\Pi$ under a probability distribution $\mu$ on
the inputs $\cX \times \cY$ be
$\delta$. Let $X, Y$ denote the random variables corresponding to
Alice's and Bob's inputs respectively.
Let $T$ denote the complete transcript of 
messages sent by Alice
and Bob. Suppose $I(XY:T) \leq a$. Let $\epsilon > 0$.
Then, there is another deterministic protocol $\Pi'$
with the following properties: 
\begin{enumerate} 
\item[(a)] The communication cost of $\Pi'$
is at most 
$\frac{2k(a+1)}{\epsilon^2} + \frac{2k}{\epsilon}$ bits;
\item[(b)] The distributional  error  of $\Pi'$ 
under $\mu$ is at most $\delta + 2\epsilon$.
\end{enumerate}
\label{thm:multicompress}
\end{theorem}
\begin{proof} 
The proof proceeds by defining
a series of intermediate $k$-round protocols
$\Pi'_k, \Pi'_{k-1}, \ldots, \Pi'_1$. 
$\Pi'_i$ is obtained from $\Pi'_{i+1}$
by compressing the message of the $i$th round. Thus, we first compress
the $k$th message, then the $(k-1)$th message, and so on. Each
message compression step introduces an additional additive error
of at most $\epsilon/k$ for every input $(x,y)$. Protocol $\Pi'_i$
uses private coins for the first $i-1$ rounds, and 
public coins for rounds $i$ to $k$. 
In fact, $\Pi'_i$ behaves the same as $\Pi$ for the first $i-1$ rounds.
Let $\Pi'_{k+1}$ denote the original protocol $\Pi$.

We now describe the construction of $\Pi'_i$ from $\Pi'_{i+1}$.
Suppose the $i$th message in $\Pi'_{i+1}$ is sent by Alice. 
Let $M$ denote the
random variable corresponding to the first $i$ messages in $\Pi'_{i+1}$.
$M$ can be expressed as $(M_1, M_2)$, where $M_2$ 
represents the random variable
corresponding to the $i$th message and $M_1$ 
represents the random variable
corresponding to the initial $i-1$ messages. 
From Fact~\ref{fact:additivity} (note that the distributions below
are as in protocol $\Pi'_{i+1}$ with the input distributed according
to $\mu$), 
\[
I(XY:M)  =  I(XY:M_1) + \E_{M_1}[I((XY:M_2) \mid M_1=m_1)] 
	 =   I(XY:M_1) + \E_{M_1XY}[S(M_2^{xym_1} \| M_2^{m_1})]
\]
where $M_2^{xym_1}$ denotes the distribution of $M_2$ when $(X,Y)=(x,y)$
and $M_1=m_1$, and $M_2^{m_1}$ denotes the distribution of 
$M_2$ when $M_1= m_1$. Note that the distribution of 
$M_2^{xym_1}$ is independent of $y$, as $\Pi'_{i+1}$ is private coin
up to the $i$th round.
Define $a_i \defeq \E_{M_{1}XY}[S(M_2^{xym_1} \| M_2^{m_1})]$.

Protocol $\Pi'_i$ behaves the same as $\Pi'_{i+1}$ for the first
$i-1$ rounds; hence $\Pi'_i$ behaves the same as $\Pi$ for the
first $i-1$ rounds. In particular, it is private coin for the
first $i-1$ rounds.
Alice generates the $i$th message of $\Pi'_i$
using a fresh public coin $C_i$ as follows: 
For each distribution $M_2^{m_1}$, $m_1$
ranging over all possible initial $i - 1$ messages, 
$C_i$ stores an infinite sequence 
$\bGamma^{m_1} \defeq \tuple{\gamma^{m_1}_j}_{j \in \nat}$,
where $(\gamma^{m_1}_j: j \in \nat)$ are chosen independently from 
distribution $M_2^{m_1}$. 
Note that the distribution $M_2^{m_1}$ is known to
both Alice and Bob as $m_1$ is known to both of them; so both Alice
and Bob know which part of $C_i$ to `look' at 
in order to read from the infinite sequence $\bGamma^{m_1}$. Using
Lemma~\ref{lem:sampling2}, Alice generates  
the $i$th message of $\Pi'_i$
which is either $x^{m_1}_j$ for some $j$, or the dummy message
$0$. The probability of generating $0$ is less than or equal to
$\frac{\epsilon}{k}$.
If Alice does not generate $0$, her message lies in a set
$\Good_{xm_1}$ which has probability at least 
$1 - \frac{\epsilon}{k}$ in the distribution $M_2^{x y m_1}$. 
The probability of a message $m_2 \in \Good_{xm_1}$ being
generated is exactly the same as the probability of $m_2$ in
$M_2^{x y m_1}$. The expected value of $j$ is 
$2^{k(S(M_2^{xym_1} \| M_2^{m_1})+1)/\epsilon}$. Actually,
Alice just sends the value of $j$ or the dummy message
$0$ to Bob, using a prefix free encoding, 
as the $i$th message of $\Pi'_i$. 
After Alice sends off the $i$th message, $\Pi'_i$ behaves the same
as $\Pi'_{i+1}$ for rounds $i + 1$ to $k$. In particular, the 
coin $C_i$ is not `used' for rounds $i + 1$ to $k$; instead, the 
public coins of $\Pi'_{i+1}$  are `used' henceforth.

By the concavity of the
logarithm function, the expected length of the $i$th message of
$\Pi'_i$ is at most
$2k\epsilon^{-1}(S(M_2^{xym_1} \| M_2^{m_1})+1)+ 2$ bits for
each $(x, y, m_1)$ (The multiplicative and additive factors of $2$
are there to take care of the prefix-free encoding).
Also in $\Pi'_i$, for each $(x, y, m_1)$,
the expected length 
(averaged over the public coins of $\Pi'_i$,
which in particular include $C_i$ and the public coins of $\Pi'_{i+1}$)
of the $(i+1)$th to $k$th messages does not
increase as compared to the expected length
(averaged over the public coins of $\Pi'_{i+1}$)
of the $(i+1)$th to $k$th messages in $\Pi'_{i+1}$. 
This is because in the 
$i$th round of $\Pi'_i$, the probability of any non-dummy message 
does not increase as compared to that in $\Pi'_{i+1}$, and if the 
dummy message $0$ is sent in 
the $i$th round $\Pi'_i$ aborts immediately.
For the same reason, the increase in the error from $\Pi'_{i+1}$ to
$\Pi'_i$ is at most an additive term of
$\frac{\epsilon}{k}$ for each $(x, y, m_1)$. Thus
the expected length, averaged over
the inputs and public and private coin tosses,
of the $i$th message in $\Pi'_i$ is at most 
$2k\epsilon^{-1}(a_i+1) + 2$ bits. 
Also, the average error
of $\Pi'_i$ under input distribution $\mu$ increases by at most
an additive term of $\frac{\epsilon}{k}$.

By Fact~\ref{fact:additivity},
$\sum_{i=i}^{k} a_i = I(XY:T) \leq a$, where $I(XY:T)$ is the
mutual information in
the original protocol $\Pi$. This is because the quantity
$\E_{M_{1}XY} [S(M_2^{xym_1} \| M_2^{m_1})]$ is the same irrespective
of whether it is calculated for protocol $\Pi$ or protocol
$\Pi'_{i+1}$, as $\Pi'_{i+1}$
behaves the same as $\Pi$ for the first $i$ rounds.
Doing the above `compression'  procedure $k$ times gives us
a public coin protocol $\Pi'_1$ such that the 
expected communication cost (averaged over the inputs as well as all
the public coins of $\Pi'_1$) of $\Pi'_1$ is at most 
$2k\epsilon^{-1}(a+1) + 2k$, 
and the average error of $\Pi'_1$ under input distribution $\mu$ is 
at most $\delta + \epsilon$. 
By restricting the maximum communication to 
$2k\epsilon^{-2}(a+1) + 2k\epsilon^{-1}$ bits and applying
Markov's inequality, we get a public coin protocol 
$\Pi''$ from $\Pi'_1$ which has average error under input 
distribution $\mu$ at most 
$\delta + 2 \epsilon$. 
By setting the public coin tosses to a suitable value, we get a
deterministic protocol $\Pi'$ from $\Pi''$ 
where the maximum communication is at most 
$2k\epsilon^{-2}(a+1) + 2k\epsilon^{-1}$ bits, and
the distributional error under $\mu$ is at most $\delta + 2\epsilon$.
\end{proof}

\begin{corollary} 
Let $f: \cX \times \cY \rightarrow \cZ$ be a function. 
Let $\mu$ be a product distribution on
the inputs $\cX \times \cY$. Let $\delta, \epsilon > 0$. Then,
\( \textstyle
IC^{k}_{\mu,\delta} (f) \geq
\frac{\epsilon^2}{2k} \cdot C^{k}_{\mu,\delta+2\epsilon}(f) - 2.
\) 
\label{cor:ICgeqC}
\end{corollary}

\begin{theorem}[Direct sum, $k$-round] 
Let $m, k$ be positive integers, and $\epsilon, \delta > 0$.
Let $f: \cX \times \cY \rightarrow {\cal Z}$ be a function. Then,
\( \textstyle 
R^{k}_{\delta}(f^m) \geq  m \cdot \sup_{\mu, \kappa} 
\left(\frac{\epsilon^2}{2k} \cdot 
      C^{k}_{\mu, \delta+2\epsilon}(f) - 2 - H(\kappa)
\right),
\)
where the supremum is over all
probability distributions $\mu$ on $\cX \times \cY$ and
partitions $\kappa$ of $\mu$.
\label{thm:directsummult}
\end{theorem}
\begin{proof} 
Immediate from
Fact~\ref{fact:CgeqIC}, Fact~\ref{fact:addinfo} and  
Corollary~\ref{cor:ICgeqC}.
\end{proof} 

\begin{corollary}
Let $m, k$ be positive integers, and $\epsilon, \delta > 0$.
Let $f: \cX \times \cY \rightarrow {\cal Z}$ be a function. Then,
\( \textstyle 
R^{k}_{\delta}(f^m) \geq  m \cdot 
\left(\frac{\epsilon^2}{2k} \cdot 
      C^{k}_{[\,], \delta+2\epsilon}(f) - 2
\right).
\)
\label{cor:directsummult}
\end{corollary}

\paragraph{Remarks:} 
\ \\
1.\ Note that all the results in this section hold even when $f$ is a
relation. \\ 2.\ The above corollary implies that the direct sum
property holds for constant round protocols for the pointer jumping
problem with the `wrong' player starting (the bit version, the full
pointer version and the tree version), since the product
distributional complexity (in fact, for the uniform distribution) of
pointer jumping is the same as its randomised
complexity~\cite{nisan:ptr,ponzio:ptr}.

\section {Impossibility of quantum compression} 
\label{sec:quantumimposs}
In this section, we show that the information cost based message
compression approach does not work in the quantum setting. 
We first need some preliminary definitions and lemmas. 

\begin{lemma}
\label{lem:discrete}
Fix positive integers $d, m$ and real
$\epsilon > 0$. Then there is a set
${\cal S}$ of at most $d$-dimensional subspaces of $\C^m$ such that
\begin{enumerate}
\item[(a)] 
$|{\cal S}| \leq \left(\frac{8\sqrt{d}}{\epsilon}\right)^{2md}$.
\item[(b)] For all $d$-dimensional subspaces $W$ of $\C^m$, there is
an at most $d$-dimensional subspace $\widehat{W} \in {\cal S}$ such that
$\Delta(W, \widehat{W}) \leq \epsilon$.
\end{enumerate}
\end{lemma}
\begin{proof}
Let ${\cal N}$ be a $\delta$-dense subset of $\bU_{m,1}$
satisfying Fact~\ref{fact:net}.
For a unit vector $v \in \C^m$, let 
$\widetilde{v}$ denote the vector in ${\cal N}$ closest to it
(ties are broken arbitrarily).
Let $W$ be a subspace of $\C^m$ of dimension $d$. 
Let $w = \sum_{i=1}^{d} \alpha_i w_i$ be a unit vector in $W$, where 
$\{w_1, \ldots, w_d\}$ is an orthonormal basis for $W$ and 
$\sum_{i=1}^{d} |\alpha_i|^2 = 1$. 
Define $w' \defeq \sum_{i=1}^{d} \alpha_i \widetilde{w}_i$ and
$\widehat{w} \defeq \frac{w'}{\norm{w'}}$ if $w' \neq 0$, 
$\widehat{w} \defeq 0$ if $w' = 0$. 
It is now easy to verify the following. 
\begin{enumerate}
\item[(a)]
$\norm{w - w'} \leq \delta\sqrt{d}$.
\item[(b)]
$\norm{w'} \geq 1 - \delta \sqrt{d}$.
\item[(c)]
$\norm{w - \widehat{w}} \leq 2 \delta \sqrt{d}$.
\end{enumerate}

Choose $\delta \defeq \frac{\epsilon}{2 \sqrt{d}}$. 
Define $\widehat{W}$ to be the subspace spanned by the set
$\{\widetilde{w_1}, \ldots, \widetilde{w_d}\}$. 
$\dim(\widehat{W}) \leq d$.
By Fact~\ref{fact:povmtrdist} and (c) above, 
$\Delta(W, \widehat{W}) \leq \epsilon$. 
Define 
${\cal S} \defeq \{\widehat{W}: W~\mbox{{\rm subspace \ of}}~\C^m~ 
                                {\rm of \ dimension}~d\}$.
${\cal S}$ satisfies part~(b) of the present lemma.
Also $|{\cal S}| \leq (4/\delta)^{2md} = (8 \sqrt{d}/\epsilon)^{2md}$,
thus proving part~(a) of the present lemma.
\end{proof}

We next prove the following two propositions using 
Fact~\ref{fact:conc}.
\begin{proposition}
\label{prop:projection1}
Let $m, d, l$ be positive integers such that $d < \sqrt{\frac{m}{l}}$
and $l < \frac{m}{20}$.
Let $V$ be a fixed subspace of $\C^m$ of dimension $m/l$. Let $P$
be the orthogonal projection operator on $V$. Let $(w, w')$ be an
independently chosen random pair of unit vectors from $\C^m$.
Then, 
\begin{enumerate}
\item[(a)]
$\Pr\left[|\braket{w}{w'}| \geq \frac{1}{5 d^2} \right] \leq
2 \exp\left(-\frac{m}{100 d^4}\right)$,
\item[(b)]
$\Pr\left[\norm{Px} \geq \frac{2}{\sqrt{l}} \right] \leq 
2 \exp \left(-\frac{m}{4l}\right), x = w, w'$,
\item[(c)]
$\Pr\left[|\unibraket{w}{P}{w'}| \geq \frac{4}{5 d^2 l} \right] \leq
6 \exp\left(-\frac{m}{100 d^4 l}\right)$.
\end{enumerate}
\end{proposition}
\begin{proof}
To prove the first inequality, we can assume by the 
$\bU(m)$-invariance of $\bmu_{m,1}$ that $w' = e_1$.
The map $w \mapsto |\braket{w}{e_1}|$ is $1$-Lipschitz, with 
expectation at most $\frac{1}{\sqrt{m}}$ by $\bU(m)$-symmetry and
using convexity of the square function. By Fact~\ref{fact:conc},
\[
\Pr\left[|\braket{w}{w'}| \geq \frac{1}{5 d^2} \right] \leq
\Pr\left[|\braket{w}{w'}| > 1/\sqrt{m} + 12/\sqrt{2m} 
         + \frac{1}{10 d^2} \right] \leq
2 \exp\left(-\frac{m}{100 d^4}\right),
\]
proving part~(a) of the present proposition.

The argument for the second inequality is similar. 
By $\bU(m)$-symmetry and using convexity of the square function,
$\E[\norm{Pw}] = \E[\norm{Pw'}] \leq \frac{1}{\sqrt{l}}$.
Since the map $w \mapsto \norm{Pw}$ is $1$-Lipschitz, by 
Fact~\ref{fact:conc} we get that
\[
\Pr\left[\norm{Px} \geq \frac{2}{\sqrt{l}}\right] \leq 
\Pr\left[\norm{Px} > \frac{1}{\sqrt{l}} + \frac{12}{\sqrt{2m}} 
                     + \frac{1}{2 \sqrt{l}}
   \right] \leq 
2 \exp \left(-\frac{m}{4l}\right), x = w, w',
\]
proving part~(b) of the present proposition.

We now prove part~(c) of the present proposition.
Let $\widehat{w} \defeq \frac{Pw}{\norm{Pw}}$ and
$\widehat{w'} \defeq \frac{Pw'}{\norm{Pw'}}$ (note that 
$\norm{Pw} = 0$ and $\norm{Pw'} = 0$ are each zero probability events).
By Fact~\ref{fact:symmetry}, $\widehat{w}, \widehat{w'}$ are 
random independently chosen unit
vectors in $V$. By the argument used in the proof of 
part~(a) of the present proposition, we get that
\[
\Pr\left[|\braket{\widehat{w}}{\widehat{w'}}| \geq \frac{1}{5 d^2} 
   \right] \leq
2 \exp\left(-\frac{m}{100 l d^4}\right).
\]
Now,
\[
\Pr\left[|\braket{Pw}{Pw'}| \geq \frac{4}{5 d^2 l} \right]  \leq 
2\exp\left(-\frac{m}{100 d^4 l}\right)
+ 4\exp\left(-\frac{m}{4l}\right) \leq 
6 \exp\left(-\frac{m}{100 d^4 l}\right),
\]
proving part~(c) of the present proposition.
\end{proof}

\begin{proposition}
\label{prop:projection2}
Let $m, d, l$ be positive integers such that $d < \sqrt{\frac{m}{l}}$
and $l < \frac{m}{20}$.
Let $V$ be a fixed subspace of $\C^m$ of dimension $m/l$. Let $P$
be the orthogonal projection operator on $V$. Let $(w, w')$ be a 
random pair of orthonormal vectors from $\C^m$.
Then, 
\[
\Pr\left[|\unibraket{w}{P}{w'}| \geq \frac{2}{d^2 l} \right] \leq
10 \exp\left(-\frac{m}{100 d^4 l}\right).
\]
\end{proposition}
\begin{proof}
By Fact~\ref{fact:symmetry}, to generate a random pair of orthonormal 
vectors $(w, w')$ from $\C^m$ we can do as follows: First generate
unit vectors $x, y \in \C^m$ randomly and independently, let
$w'' \defeq y - \braket{x}{y} x$, and set $w \defeq x$ and
$w' \defeq \frac{w''}{\norm{w''}}$. Now (note that $\Pr[w'' = 0] = 0$),
\[
|\unibraket{w}{P}{w'} = \frac{|\unibraket{w}{P}{w''}|}{\norm{w''}}
\leq \frac{|\unibraket{x}{P}{y} + |\braket{x}{y}| 
           \unibraket{x}{P}{x}}
          {1 - |\braket{x}{y}|}.
\]
By Proposition~\ref{prop:projection1} we see that,
\begin{eqnarray*}
\Pr\left[|\unibraket{w}{P}{w'}| \geq \frac{2}{d^2 l} \right]
& \leq & 
\Pr\left[|\unibraket{w}{P}{w'}| \geq 
\frac{4/(5 d^2 l) + (1/(5d^2)) \cdot (4/l)}{1 - (1/(5d^2))} \right] \\
& \leq & 6 \exp\left(-\frac{m}{100 d^4 l}\right) 
         + 2 \exp\left(-\frac{m}{100 d^4}\right)
         + 2 \exp\left(-\frac{m}{4 l}\right) \\
& \leq & 10 \exp\left(-\frac{m}{100 d^4 l}\right),
\end{eqnarray*}
proving the present proposition.
\end{proof}

\begin{lemma}
\label{lem:subspace}
Let $m, d, l$ be positive integers such that
$200 d^4 l \ln(20 d^2) < m$.
Let $V$ be a fixed subspace of $\C^m$ of dimension $m/l$. Let $P$
be the orthogonal projection operator on $V$. Let $W$ be a random
subspace of $\C^m$ of dimension $d$. Then,
\[
\Pr[\exists w \in W, \, \norm{w}=1 {\rm ~~and~~} 
    |\unibraket{w}{P}{w}| \geq 6/l] \leq
\exp\left(-\frac{m}{200 d^4 l}\right).
\]
\end{lemma}
\begin{proof}
Let $(w_1, \ldots, w_d)$ be a randomly chosen ordered orthonormal
set of size $d$ in $\C^m$, and let 
$W \defeq \linspan (w_1, \ldots, w_d)$. 
By Fact~\ref{fact:symmetry}, each $w_i$ is a random unit vector
of $\C^m$ and each $(w_i, w_j)$, $i \neq j$ is a random pair
of orthonormal vectors of $\C^m$. 
By Propositions~\ref{prop:projection1} and \ref{prop:projection2},
we have with probability at least
$1 - 2 d \exp\left(-\frac{m}{4 l}\right) -
10 d^2 \exp\left(-\frac{m}{100 d^4 l}\right)$,
\[
\forall i, \unibraket{w_i}{P}{w_i} < \frac{4}{l} 
~~{\rm and}~~
\forall i, j, i \neq j, |\unibraket{w_i}{P}{w_j}| < \frac{2}{d^2 l}.
\]
We show that whenever this happens $|\unibraket{w}{P}{w}| \leq 6/l$
for all $w \in W$, $\norm{w} = 1$. 
Let $w \defeq \sum_{i=1}^d \alpha_i w_i$, where 
$\sum_{i=1}^d |\alpha_i|^2 = 1$. Then,
\begin{eqnarray*}
|\unibraket{w}{P}{w}| 
&   =  & \left| \sum_{i,j} \alpha_i^\ast \alpha_j 
                           \unibraket{w_i}{P}{w_j} \right| \\
& \leq & \sum_i |\alpha_i|^2 |\unibraket{w_i}{P}{w_i}| +
         \sum_{i,j: i \neq j} |\alpha_i^\ast \alpha_j|
                              |\unibraket{w_i}{P}{w_j}|    \\
&   <  & \frac{4}{l} + d^2 \cdot \frac{2}{d^2 l} \\
&   =  & \frac{6}{l}.
\end{eqnarray*}
Thus, 
\begin{eqnarray*}
\Pr[\exists w \in W, \, \norm{w}=1 {\rm ~~and~~} 
    |\unibraket{w}{P}{w}| \geq 6/l] 
& \leq & 2 d \exp\left(-\frac{m}{4 l}\right) 
         + 10 d^2 \exp\left(-\frac{m}{100 d^4 l}\right) \\
& \leq & \exp\left(-\frac{m}{200 d^4 l}\right),
\end{eqnarray*}
completing the proof of the present lemma.
\end{proof}

We can now prove the following `incompressibility' theorem about
(mixed) state compression in the quantum setting.
\begin{theorem}[Quantum incompressibility]
\label{thm:incompress}
Let $m, d, n$ be positive integers and $k$ a positive real number
such that $k > 7$, $d > 160^2$,
$1600 d^4 k 2^k \ln(20 d^2) < m$ and
$3200 2^{2k} d^5 \ln d < n$.
Let the underlying Hilbert space be $\C^m$. There exist $n$
states $\rho_l$ and $n$ orthogonal projections $M_l$,
$1 \leq l \leq n$ such that
\begin{enumerate}
\item[(a)] $\forall l \, \Tr M_l \rho_l = 1$.
\item[(b)] $\rho \defeq \frac{1}{n} \cdot \sum_l \rho_l 
                    =   \frac{1}{m} \cdot I$,
           where $I$ is the identity operator on $\C^m$.
\item[(c)] $\forall l \, S(\rho_l \| \rho) = k$.
\item[(d)] For all subspaces $W$ of dimension $d$, 
           $|\{M_l: M_l(W) \leq 1/10\}| \geq n/4$.
\end{enumerate}
\end{theorem}
\begin{proof}
In the proof, we will index the $n$ states $\rho_l$, $1 \leq l \leq n$
as $\rho_{ij}$, $1 \leq i \leq \frac{n}{2^k}$, $1 \leq j \leq 2^k$.
We will also index the $n$ orthogonal projections $M_l$ as
$M_{ij}$.
For $1 \leq i \leq \frac{n}{2^k}$, choose 
${\cal B}^i = (\ket{b_1^i}, \ldots, \ket{b_m^i})$ to be a random
ordered orthonormal basis of $\C^m$. 
${\cal B}^i$ is chosen independently
of ${\cal B}^{i'}$, $i' \neq i$. Partition the sequence ${\cal B}^i$
into $2^k$ equal parts; call these parts ${\cal B}^{ij}$,
$1 \leq j \leq 2^k$. 
Define $\rho_{ij} \defeq \frac{2^k}{m} \cdot 
\sum_{v \in {\cal B}^{ij}} \ketbra{v}$. 
Define $M_{ij} \defeq \sum_{v \in {\cal B}^{ij}} \ketbra{v}$. 
Define $V_{ij} \defeq \linspan (v: v \in {\cal B}^{ij})$. 
$V_{ij}$ is the support of $\rho_{ij}$.
It is easy to see that $\rho_{ij}, M_{ij}$ satisfy 
parts~(a), (b) and (c) of the present theorem. 

To prove part~(d), we reason as follows. Let $W$ be a fixed subspace
of $\C^m$ of dimension $d$. 
Let $P_{ij}$ denote the orthogonal projection operator
onto $V_{ij}$. 
By the $\bU(m)$-invariance of the distribution $\bmu_{m,d}$
and from Lemma~\ref{lem:subspace}, for each $i, j$,
\[
\Pr\left[\exists w \in W, \, \norm{w}=1 {\rm ~~and~~} 
    |\unibraket{w}{P_{ij}}{w}| \geq \frac{6}{2^k}
   \right] \leq
\exp\left(-\frac{m}{200 \cdot 2^k d^4}\right),
\]
where the probability is over the random choice of the bases
${\cal B}^i$, $1 \leq i \leq \frac{n}{2^k}$.
Define the set 
\[
\Bad \defeq \{i \in [n/2^k]: \exists j \in [2^k], 
                             M_{ij}(W) \geq \frac{6}{2^k}\}.
\]
Hence for a fixed $i \in \left[\frac{n}{2^k}\right]$, 
\[
\Pr[i \in \Bad] 
\leq 2^k \exp\left(-\frac{m}{200 \cdot 2^k d^4}\right) 
\leq \exp\left(-\frac{m}{400 \cdot 2^k d^4}\right).
\]
Since the events $i \in Bad$ are independent, 
\[
\Pr\left[|\Bad| \geq \frac{3}{4} \cdot \frac{n}{2^k}\right] 
\leq {\frac{n}{2^k} \choose \frac{3 n}{4 \cdot 2^k}}
     \exp\left(-\frac{3mn}{1600 \cdot 2^{2k} d^4}\right).
\leq \left(\frac{4e}{3}\right)^{\frac{3n}{2^{k+2}}} 
     \exp\left(-\frac{3mn}{1600 \cdot 2^{2k} d^4}\right).
\]
So, 
\[
\Pr\left[\left|\left\{M_{ij}: M_{ij}(W) \geq \frac{6}{2^k}\right\}
         \right| \geq \frac{3n}{4}
   \right] 
\leq \left(\frac{4e}{3}\right)^{\frac{3n}{2^{k+2}}} 
     \exp\left(-\frac{3mn}{1600 \cdot 2^{2k} d^4}\right).
\]
By setting $\epsilon = 1/20$ in Lemma~\ref{lem:discrete}, we get 
\begin{eqnarray*}
\lefteqn{
\Pr\left[\exists \widehat{W} \in {\cal S}, 
         \left|\left\{M_{ij}: M_{ij}(W) \geq \frac{1}{20}\right\}
	 \right| \geq \frac{3n}{4}
   \right]
} & & \\
& \leq & \left(\frac{4e}{3}\right)^{\frac{3n}{2^{k+2}}} 
	 (8 \sqrt{d} / \epsilon)^{2md}
         \exp\left(-\frac{3mn}{1600 \cdot 2^{2k} d^4}\right) \\
&   <  & 1,
\end{eqnarray*}
for the given constraints on the parameters. Again by
Lemma~\ref{lem:discrete}, we get
\begin{eqnarray*}
\lefteqn{
\Pr\left[\exists W~{\rm subspace \ of}~\C^m,  \dim(W) = d, 
         \left|\left\{M_{ij}: M_{ij}(W) \geq \frac{1}{10}\right\}
	 \right| \geq \frac{3n}{4}
   \right]
} & & \\
&   =  & \Pr\left[\exists \widehat{W} \in {\cal S}, 
                  \left|\left\{M_{ij}: 
			M_{ij}(W) \geq \frac{1}{20}\right\}
	          \right| \geq \frac{3n}{4}
            \right] \\
&   <  & 1.
\end{eqnarray*}
This completes the proof of part~(d) of the present theorem.
\end{proof}

\section{Conclusion and open problems}
\label{sec:conclusion}
In this paper, we have shown a 
compression theorem and a direct sum theorem for two party multiple
round private coin protocols. Our proofs use the notion of information
cost of a protocol. The main technical ingredient in our compression
proof is a connection between relative entropy and sampling. It
is an interesting open problem to strengthen this connection, so as
to obtain better lower bounds for the direct sum problem for
multiple round protocols. In particular, can one improve the dependence
on the number of rounds in the compression result (by information
cost based methods or otherwise)? 

We have also shown a strong negative result about the compressibility
of quantum information. Our result seems to suggest that to
tackle the direct sum problem in quantum communication, techniques
other than information cost based message compression may be
necessary. Buhrman et al.~\cite{buhrman:fingerprint} have shown that
the bounded error simultaneous 
quantum complexity of $\EQ_n$ is $\theta(\log n)$, as opposed to
$\theta(\sqrt{n})$ in the classical 
setting~\cite{newman:eq,babai:eq}.
An interesting open problem is whether the direct sum property holds
for simultaneous quantum protocols for equality. 

\section*{Acknowledgements}
We thank Ravi Kannan and Sandeep Juneja 
for helpful discussions, and Siddhartha Bhattacharya for
enlightening us about unitarily invariant measures on homogeneous
spaces. We also thank the anonymous referees
for their comments on the conference version of this paper, 
which helped us to improve the presentation of the paper.

\bibliography{directsum}

\end{document}